\documentclass[sigconf]{acmart}

\AtBeginDocument{%
  \providecommand\BibTeX{{%
    \normalfont B\kern-0.5em{\scshape i\kern-0.25em b}\kern-0.8em\TeX}}}

\copyrightyear{2022} 
\acmYear{2022} 
\setcopyright{acmlicensed}\acmConference[W4A'22]{19th Web for All Conference}{April 25--26, 2022}{Lyon, France}
\acmBooktitle{19th Web for All Conference (W4A'22), April 25--26, 2022, Lyon, France}
\acmPrice{15.00}
\acmDOI{10.1145/3493612.3520448}
\acmISBN{978-1-4503-9170-2/22/04}

\usepackage{xspace}

\newcommand{\eg}{\textit{e.g.}\@\xspace}
\newcommand{\ie}{\textit{i.e.}\@\xspace}
\newcommand{\etal}{\textit{et al.}\@\xspace}

\usepackage[labelformat=simple]{subcaption} 

\usepackage{makecell}
\usepackage{multirow}
\usepackage{enumitem}
\usepackage[export]{adjustbox}
\usepackage{colortbl}
\definecolor{LightGray}{gray}{0.9}

\begin{document}


\title[From the Lab to People’s Home: Lessons from Accessing Blind Participants' Interactions via Smart Glasses]{From the Lab to People’s Home: Lessons from Accessing Blind Participants' Interactions via Smart Glasses in Remote Studies}
\author{Kyungjun Lee}
\authornote{Both authors contributed equally to this research work.}
\email{kjlee@cs.umd.edu}
\orcid{0000-0001-8556-9113}
\affiliation{%
  \institution{University of Maryland}
  \city{College Park}
  \state{MD}
  \country{USA}
  \postcode{20742}
}

\author{Jonggi Hong}
\authornotemark[1]
\email{jhong@ski.org}
\affiliation{%
  \institution{Smith-Kettlewell Eye Research Institute}
  \city{San Francisco}
  \state{CA}
  \country{USA}
  \postcode{94115}
}

\author{Ebrima Jarjue}
\email{ebjarjue@terpmail.umd.edu}
\affiliation{%
  \institution{University of Maryland}
  \city{College Park}
  \state{MD}
  \country{USA}
  \postcode{20742}
}

\author{Ernest Essuah Mensah}
\email{eessuahmensah@apple.com}
\affiliation{%
  \institution{Apple Inc.}
  \city{Cupertino}
  \state{CA}
  \country{USA}
}

\author{Hernisa Kacorri}
\email{hernisa@umd.edu}
\affiliation{%
 \institution{University of Maryland}
 \city{College Park}
 \state{MD}
 \country{USA}
}

\renewcommand{\shortauthors}{Lee and Hong, et al.}

\begin{abstract} 
Researchers have adopted remote methods, such as online surveys and video conferencing, to overcome challenges in conducting in-person usability testing, such as participation, user representation, and safety. However, remote user evaluation on hardware testbeds is limited, especially for blind participants, as such methods restrict access to observations of user interactions. We employ smart glasses in usability testing with blind people and share our lessons from a case study conducted in blind participants' homes ($N$=12), where the experimenter can access participants' activities via dual video conferencing: a third-person view via a laptop camera and a first-person view via smart glasses worn by the participant. We show that smart glasses hold potential for observing participants' interactions with smartphone testbeds remotely; on average 58.7\% of the interactions were fully captured via the first-person view compared to 3.7\% via the third-person. However, this gain is not uniform across participants as it is susceptible to head movements orienting the ear towards a sound source, which highlights the need for a more inclusive camera form factor. We also share our lessons learned when it comes to dealing with lack of screen reader support in smart glasses, a rapidly draining battery, and Internet connectivity in remote studies with blind participants.
\end{abstract}

\ccsdesc[500]{Human-centered computing~Accessibility design and evaluation methods}

\keywords{remote method, user study, blind people, smart glasses}

\maketitle

\begin{figure}[t]
    \centering
    \includegraphics[width=0.48\textwidth]{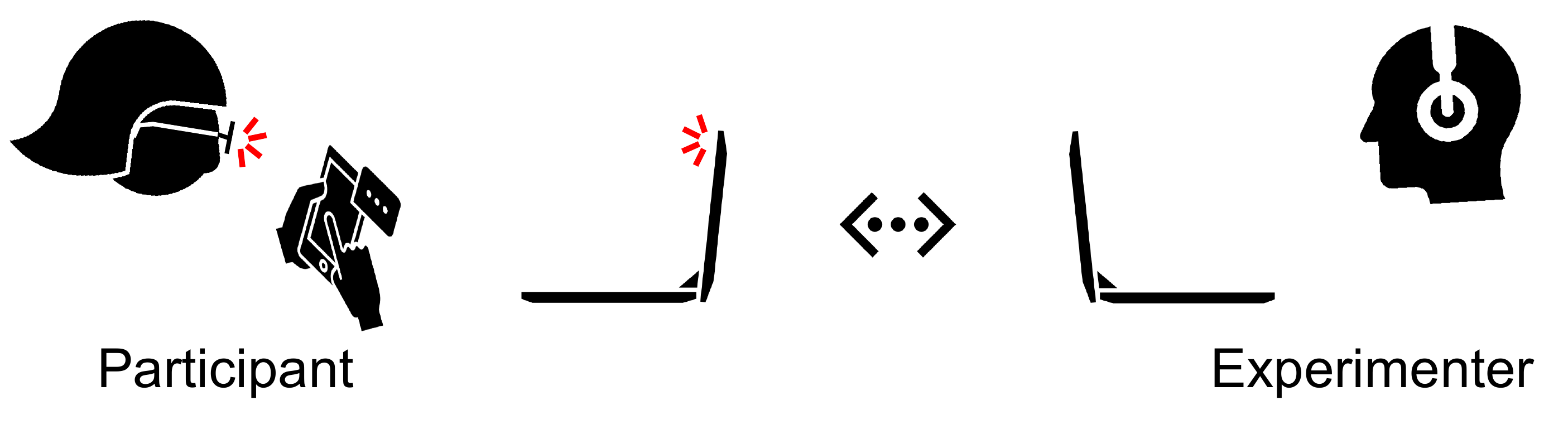}
    \Description[A diagram showing a blind participant on the left and a sighted experimenter on the right are remotely communicating via video conferencing.]{On the left, a blind participant interacts with a testbed on a smartphone while wearing smart glasses. The camera of a laptop and smart glasses capture the participant's activities and share them with the experimenter via video conferencing. The experimenter on the right monitors the blind participant's activities captured by the laptop camera and smart glasses on his laptop and communicates with the participant when necessary via video conferencing.}
    \caption{Dual video conferencing in our remote study design. While interacting with a testbed on a smartphone, a blind participant wearing smart glasses communicates with the experimenter through dual video conferencing. Two video streams are being sent to the experimenter: one from the participant's smart glasses and the other from a laptop camera facing the participant on the same Zoom call.}
    \label{fig:remote_study_design}
\end{figure}

\section{Introduction}
Approaches and tools for supporting remote user studies play an important role in human-computer interaction as they can help overcome some of the limitations of in-person studies, such as low recruitment of representative participants~\cite{petrie2006remote}. With the recent COVID-19 outbreak~\cite{whocovid19}, remote methods have received more attention both to comply with government guidance and to ensure the safety and comfortableness of all involved. In response, the accessibility community has employed online surveys~\cite{gonccalves2020playing, lee2021accesscomics, leporini2021distance}, audio calls~\cite{gleason2020making, lee2020tableview, saha2020understanding, ahmetovic2021musa, jain2021smartphone}, and video conferencing~\cite{troncoso2020aiguide, schaadhardt2021understanding} to relieve the safety concerns of participants with visual impairments related to transportation and face-to-face interactions~\cite{rosenblum2020flatten}.

Cameras available on laptops or attached to desktops with video conferencing tools, such as Google Meet~\cite{meet}, Microsoft Teams~\cite{teams}, and Zoom~\cite{zoom}, have enabled experimenters to observe participants' interactions~\cite{kohlrausch2005audio}, which can be supplementary to traditional remote methods for usability testing~\cite{simon2021remote}.
Similar setups have also been employed in remote usability testing with blind participants~\cite{troncoso2020aiguide}. 
However, the visual information from such stationary camera may be limited in that it only captures participants' activities from a single third-person viewpoint, which may not show participants' real-time interactions with the user interface of smartphone or smartwatch testbeds (\ie, study prototypes). Screen sharing from those devices can allow such observations, but it is not trivial to participants. This is especially the case for blind participants due to its inaccessibility~\cite{miao2016contrasting}.
In this paper, we explore ways to overcome these challenges associated with remote observations of blind participants' interactions via video conferencing with smart glasses.

\textbf{Why smart glasses in remote user studies with the blind.} Typically, during in-person studies where blind participants are asked to interact with a testbed or a working application, experimenters employ a plethora of methods for collecting data including but not limited to: (i) video recordings via a carefully framed static camera placed on markings in front or to the left/right of the participant to ensure systematic recordings across participants; (ii) field notes based on their own observations; and (iii) application logs capturing fine-grained interaction data. Experimenters are also strategically seated close to the participant for observation, guidance, and troubleshooting as the testbed or prototype may not be fully functional. Replicating this experimental setup in blind participants' homes can be challenging. Laptops with video conferencing are usually placed in front of the participant. Thus, the third-person viewpoint may not capture the necessary context for the interactions. On the other hand, smart glasses worn by the participant and equipped with camera and video conferencing functionalities can provide access to the first-person viewpoint, which may \textit{increase remote access} and \textit{support communication}. 

We explore these premises with the following research questions: RQ1 \textit{``What do smart glasses worn by blind people capture during a remote user study?''} and RQ2 \textit{``How can this be leveraged to increase access to their interactions with the system being evaluated and better support experimenter-participant communication?''}
Specifically, we investigate the feasibility and challenges of using smart glasses in video conferencing for a user study that, due to the pandemic, had to move from a lab to blind participants' houses.
To that end, we devise a remote experimental setup and protocol for observing blind participants' interactions with smartphone testbeds and object stimuli. 
As depicted in Figure~\ref{fig:remote_study_design}, the experimenter can access participants' activities via dual video conferencing where the third-person view is captured by a laptop camera and the first-person view from smart glasses worn by the participant. Both devices are connected to the same video conference call, but, to prevent audio echo, only the laptop sound is used. After multiple iterations of piloting among blind and sighted researchers, the protocol is deployed in the homes of 12 blind participants serving as a case study. Through peer debriefing and fine-grained analysis of study recordings, we explore the feasibility of this approach in terms of \textbf{access} (\ie, by looking at the visual information captured by the smart glasses versus the laptop), \textbf{support} (\ie, by looking at the experimenter-participant communication), and \textbf{logistics} (\ie, by reflecting on our experiences with handling delivery and troubleshooting).

We find that the video conferencing from the smart glasses provided the experimenter with real-time remote access to blind users’ interactions. On average 58.7\% of the video frames from the smart glasses fully captured the interactions. This was strikingly high compared to those captured by the laptop camera (3.7\%). More so, the experimenter could leverage partially visible interactions on the smart glasses and triangulate them with audio cues such as the screen reader output on the testbed. We also find that the smart glasses supported the experimenter in providing guidance and making observations in an unobtrusive way. Interruptions related to camera aiming were minimal and only occurred for a few participants, usually during onboarding.

While these findings are promising, what can be captured by smart glasses is not uniform among participants. More interactions are captured for those who became blind later in life compared to those who were born blind and tended to ``turn their head to orient their ear towards a sound source''~\cite{wanet1985processing, martinez1977informations, lewald2002opposing}. It highlights the importance of the camera field of view for the inclusion of blind users. The lack of screen reader support on smart glasses, a rapidly draining battery, and a dependency on Internet connection called for a nearby presence of the experimenter. Even workarounds like long power cables and portable chargers can fail as cable disconnection can be inaccessible for blind participants to spot. This unpredictability suggests that smart glasses should be used as a complement to a typical laptop video call or phone call setup rather than a substitute.

This paper's contributions are: (i) a novel approach for facilitating first-person access to participants' interactions in remote studies via smart glasses; (ii) empirical results with blind participants on the effect of the camera field of view and sound source on what is being captured by their head-worn cameras; and (iii) insights to the challenges and logistics involved in conducting remote studies and employing state-of-the-art smart glasses in them.


\section{Related Work}
Our work draws upon existing literature in accessibility for conducting remote user studies with a focus on studies that involve participants with visual impairments. To provide broader context for our approach, we discuss prior work in remote sighted guidance and assistive smart glasses for this population.


\subsection{Remote User Studies in Accessibility}

\begin{table*}[t]
    \small
    \centering
    \caption{Approaches in remote studies with blind and low vision participants during the pandemic (2020--2021). 
    }
    \begin{tabular}{l|cccc|m{0.32\textwidth}}
        \toprule
        \textbf{Prior work} & \textbf{Online text} & \textbf{Audio call} & \textbf{Video call} & \textbf{Screenshare} & \textbf{Technology} \\
        \hline
        \rowcolor{LightGray}
        Gleason \etal, 2020~\cite{gleason2020making}        &           & $\bullet$ &           &           & laptop/phone: audio call \\
        Gonccalves \etal, 2020~\cite{gonccalves2020playing} & $\bullet$ &           &           &           & web browser: online survey \\
        \rowcolor{LightGray}
        Akter \etal, 2020~\cite{akter2020privacy}           & $\bullet$ &           &           &           & web browser: online survey \\
        Troncoso \etal, 2020~\cite{troncoso2020aiguide}&          &           & $\bullet$ &           & laptop: video call \\ 
        \rowcolor{LightGray}
        Engel \etal, 2020~\cite{engel2020travelling}        & $\bullet$ &           &           &           & web browser: online survey \\
        Lee \etal, 2020~\cite{lee2020tableview}             &           & $\bullet$ &           & $\bullet$ & laptop: audio call, screen share  \\ 
        \rowcolor{LightGray}
        Saha \etal, 2020~\cite{saha2020understanding}       &           & $\bullet$ &           &           & laptop/phone: audio call \\
        Ahmetovic \etal, 2021~\cite{ahmetovic2021musa}      &           & $\bullet$ &           &           & phone: audio call \\ 
        \rowcolor{LightGray}
        Lee \etal, 2021~\cite{lee2021accesscomics}          & $\bullet$ &           &           &           & web browser: online survey \\
        Leporini \etal, 2021~\cite{leporini2021distance}    & $\bullet$ &           &           &           & web browser: online survey  \\
        \rowcolor{LightGray}
        Siu \etal, 2021~\cite{siu2021covid19}               & $\bullet$ & $\bullet$ &           & $\bullet$ & web browser: online survey; computer: audio call, screen share  \\
        Chung \etal, 2021~\cite{chung2021improving}         & $\bullet$ & $\bullet$ &           &           & web browser: online survey; phone: audio call \\
        \rowcolor{LightGray}
        Jain \etal, 2021~\cite{jain2021smartphone}          &           & $\bullet$ &           & $\bullet$ & laptop: screen share; phone: audio call, app logging  \\ 
        Schaadhardt \etal, 2021~\cite{schaadhardt2021understanding} &   &           & $\bullet$ & $\bullet$ & laptop: video call, digital artboards \\
        \rowcolor{LightGray}
        Wang \etal, 2021~\cite{wang2021revamp}              &           & $\bullet$ &           &           & laptop: audio call \\ 
        \bottomrule
    \end{tabular}
    \label{tab:remote_study}
\end{table*}

User evaluation with a large number of participants with disabilities or older adults is often challenging for accessibility researchers~\cite{petrie2006remote}. Typically, our community involves a small number of representative users (median sample size 13~\cite{mack2021what}) in researchers' local area~\cite{sears2011representing}. In some cases, remote user study methods are adopted to overcome constraints in participants' location and time~\cite{petrie2006remote}.
Real-world deployments are also employed (\eg,~\cite{bigham2009evaluating}), though they require the development of fully accessible and functioning applications, which can be challenging in prototype evaluation. Discussions on the benefits of remote user evaluation have often revolved around reducing travel cost~\cite{schnepp2011improving} and decoupling the effects of time and space~\cite{andreasen2007what}.

After the outbreak of COVID-19~\cite{whocovid19}, many accessibility researchers switched to a remote format for their user studies and shared best practices with the community~\cite{dixon2021lessons, wood2021investigating, simon2021remote}.
With a focus on remote studies that involve people with visual impairments over the past two years (Table~\ref{tab:remote_study}), we observe that common approaches employed by researchers spanned online surveys, audio calls, and video conferencing.
The majority conducted either an online survey~\cite{gonccalves2020playing, akter2020privacy, engel2020travelling, lee2021accesscomics, leporini2021distance} or an interview via a call~\cite{gleason2020making, saha2020understanding, lee2020tableview, ahmetovic2021musa, jain2021smartphone, wang2021revamp}; few did both~\cite{siu2021covid19, chung2021improving}.
Some of them were able to evaluate their prototypes after guiding participants to install on their devices either a web browser extension~\cite{lee2020tableview, wang2021revamp} or a smartphone app~\cite{ahmetovic2021musa, jain2021smartphone}. One of them~\cite{lee2020tableview} also employed screen recording in their testbed, which was deployed on a web browser on participants' laptops. We see screen sharing in other studies that don't involve audio calls. However, they also are restricted to laptops either web browsers~\cite{siu2021covid19} or digital artboards~\cite{schaadhardt2021understanding}.
More importantly, we don't see any prior work employing this screen sharing feature on smartphone. Instead, we see researchers (\eg,~\cite{troncoso2020aiguide}) recording participants' interactions with their prototype system using a laptop camera (\ie, a camera that provides only a stationary and third-person view). While the recording from a third-person point of view may serve as a proxy for recording users' interactions from a stationary camera in an in-person user study, it has its limitations. For example, it would be difficult for experimenters to remotely observe both the user interface on the smartphone and participants' interactions with it at the same time. We believe that cameras embedded in smart glasses can play a role in filling this gap as they can provide a mobile and first-person perspective, which may or may not capture both the interface and participant's interactions with it. To our knowledge, using smart glasses for this task has not been previously explored with blind participants, whose idiosyncratic head movements could potentially affect what is being captured by the camera~\cite{lee2021accessing}.

\subsection{People with Visual Impairments, Remote Guidance, and Asssistive Smart Glasses}
Many studies have looked at the potential of providing remote sighted guidance to people with visual impairments. The context for guidance varies from having access to visual surroundings (\eg, objects~\cite{bigham2009evaluating} and people~\cite{baranski2015field}) and supporting indoor navigation~\cite{chaudary2017tele, rafian2017remote, kamikubo2020support}, to remote mobility training~\cite{dewald2013feasibility, barrett2016roam}.
For this remote guidance from sighted crowd, people with visual impairments often share their camera view either on their smartphone or smart glasses and post associated questions. Upon receiving this information, sighted people either provide an answer or inquire more information by interacting with the users with visual impairments.
In particular, our community has provided suggestions on remote guidelines that sighted people should follow to help people with visual impairments in indoor navigation~\cite{kamikubo2020support} and remote mobility training~\cite{dewald2013feasibility, barrett2016roam} when communicating via video calls.
This access paradigm has long moved from research labs to real-world applications, such as BeMyEyes~\cite{BeMyEyes}, TapTapSee~\cite{TapTapSee}, and Aira~\cite{Aira}.
Inspired by and complementary to these efforts, our work expands this concept of remote guidance from short instance interactions to longer experimenter-participant communication in remote user evaluation and explores the potential of smart glasses for this task.

The use of smart glasses by people with visual impairments is not new in the accessibility field. Typically, prior work has been interested in automatically leveraging the input from smart glasses camera to support people with visual impairments navigate indoors~\cite{fiannaca2014headlock, al2016ebsar, zhao2020effectiveness}, read text~\cite{zhao2015foresee, stearns2018design, huang2019augmented}, or detect objects~\cite{everingham1999head} and people~\cite{stearns2018automated, lee2020pedestrian, morrison2021social} without the explicit help from sighted people. 
Assuming that the field of view of the camera worn by users with visual impairments can capture the area of interest to them, all the prior work proposes assistive systems that interpret the visual input and convert it into non-visual formats, such as audio and haptic.
While many focused on assistive applications of smart glasses for people with visual impairments, Lee \etal focused on blind people's camera aiming behaviors in the context of pedestrian detection~\cite{lee2021accessing}. They observed that smart glasses worn by blind people tend to capture a passerby well in a corridor but may exclude a passerby from the camera frame when they are close to the passerby due to the behavior of focusing on the sound source (\ie, the passerby's voice in their case). 
Our discussions are more related to the later in that we also investigate how smart glasses worn by blind participants help experimenters observe the participants' activity. While the contexts differ (tracking people versus tracking the interface of the smartphone blind people may be interacting with), we anticipate that there could be similar challenges. Thus, our work could contribute to a broader understanding of smart glasses and their potential and limitations for the blind community.

\section{Method}
Our team, including four sighted researchers (R1, R2, R4 \& R5)  and one blind researcher (R3), used an iterative design process to devise a remote user study approach for the evaluation of a smartphone testbed employing two different camera form factors: a laptop camera and a camera embedded in smart glasses. 
Then, we explored the potential and limitations of this approach in a remote study with 12 blind participants, serving as a case study.



\subsection{Step 1: Exploration of Smart Glasses}

Prior to leveraging smart glasses in remote studies, two sighted researchers (R1 and R5) had extensive experience with employing this technology for in-person studies with blind participants. Specifically, in collaboration with other sighted and blind collaborators, they used a pair of Vuzix Blade smart glasses~\cite{VuzixBlade} to explore blind people's camera aiming behaviors for pedestrian detection~\cite{lee2021accessing} and understand its social acceptability~\cite{lee2020pedestrian}. 

In this initial investigation, blind participants were walking or standing, and the goal was for the smart glasses to capture any nearby pedestrians. While this scenario is different from being seated and interacting with a smartphone, it allowed us to gain the following insights that we leveraged in our approach:
\begin{itemize}
    \item Smart glasses capture the blind wearer's viewpoint.
    \item Blind people tend to appreciate the ability to aim the camera on the smart glasses without using their hands.
    \item Blind people may not feel comfortable wearing earbuds along with smart glasses as they limit access to the surroundings --- prior versions of Vuzix Blade were not equipped with speakers; more so, the form factor of many smart glasses includig Vuzix Blade conflicts with bone-conduction headphones. 
    \item Camera viewpoint is susceptible to head movements, which can be challenging as some blind people may ``turn their head to orient their ear towards a sound source''~\cite{wanet1985processing, martinez1977informations}.

\end{itemize}

Since the smart glasses tend to capture the blind wearer's viewpoint, we saw the potential of using this device in a remote evaluation study setup, especially for the purpose of observation. Vuzix Blade runs on Android and supports Zoom video conferencing~\cite{zoom}. Zoom on the smart glasses could enable the wearer to share the field of their camera view and communicate with others (\eg, experimenter) on the Zoom session through the built-in camera and mic. However, Vuzix Blade does not support TalkBack (the screen reader feature for Android), which is critical to blind users\footnote{To our knowledge, there are currently no commercially available smart glasses that support screen readers and remote video conferencing.}.  Also, although we found that the smart glasses can automatically launch Zoom at startup, sight is necessary for navigating the visual user interface on the glasses to join a specific Zoom meeting.
To work around this inaccessibility, a sighted experimenter ought to set up the Zoom session on the smart glasses before delivering them over to blind participants. This is a huge limitation, and more work is needed to investigate such inaccessibility issues on smart glasses.

\subsection{Step 2: Iterative Design via Piloting}

We explored how smart glasses can be deployed in a remote study with blind participants in the context of evaluating smartphone applications (case study). To this end, we conducted four pilot sessions where we iterated on the approach for incorporating the smart glasses in a predefined remote study protocol.  All pilot sessions were conducted remotely by R2, who is sighted, and R3, who is blind. During each session, R3 focused on clarifying, reviewing, and checking the study procedures from the blind participant's perspective. After each session, R2 and R3 met with R1, R4, and R5 to reflect. The planned case study, discussed in Section 3.3, included: an interview, a series of tasks with Testbed A, a series of tasks with Testbed B, and an open-ended questionnaire. 

The use of smart glasses was critical for capturing participants' interactions with the mobile applications serving as testbeds.
However, since the smart glasses had to be ON and set up with Zoom, our first pilot was merely to estimate the duration of receiving the equipment and conducting the initial interview, which lasted about an hour. The second pilot focused on tasks with Testbed A, which took about 48 minutes. It highlighted challenges in connecting with the WiFi in blind people's home and battery life for the smart glasses (initial version of Vizux Blade did not last more than 30 minutes). In the third pilot, we focused on the remaining tasks with testbed B and the last questionnaire, which took about 2 hours. Here, we opted for an upgraded version of Vizux Blade, which had a built-in speaker. However, its battery life still remained limited (no more than 40 minutes). For our last pilot, we shortened questionnaires and updated the study protocol to reflect our observations. We obtained a WiFi hotspot and connected all devices: laptop, smartphone, and the upgraded Vizux Blade, which was now also connected to a portable charger. The overall session, including the interview, tasks with testbeds A \& B, and open-ended questions lasted 2 hours. The cable connecting the smart glasses to the charger ended up being too short constraining R3's movements. We switched to a longer cable (1.2m).
In a nutshell, we saw that:
\begin{description}[leftmargin=0.3cm]
    \item[Wearing earphones limits participants' interactions.] Given a quiet environment during the study, participants listen to the device they interact with and the experimenter at the same time. It thus makes it more critical for smart glasses to have built-in speakers to facilitate both listening and communicating.
    \item[Video streaming drains smart glasses' battery.] Battery life for smart glasses is critical and quite susceptible to video streaming. Portable batteries work but can limit natural interactions.
    \item[Reliable and fast internet connection is critical.] Insufficient network bandwidth may cause frequent lagging and freezing issues as it can be quickly exhausted by two cameras on the laptop and smart glasses streaming videos simultaneously and testbed apps sending over photos and logs. Portable WiFi devices that provide 5G Nationwide or 4G LTE data speed may help.
    \item[Internet configurations can be challenging and inaccessible.] Connecting devices to the Internet can be a nontrivial task for many, especially when not familiar with study devices. For blind participants, it can also be inaccessible. Having a portable WiFi can also prevent participants from connecting study devices to their home Internet; devices can remember the portable WiFi access point and automatically connect to it when available.  
    \item[Smartphone screen recording may not be reliable.] In our pilot, screen recording needed to be enabled at the start of each task and stopped once completed, but some recordings stopped without a notification once the smartphone screen goes off due to screen timeout or an accidental button push.
    \item[Nearby presence may be vital for troubleshooting.] Technical difficulties and disruptions can be frustrating without support and lead to a confounding effect. Beyond step-by-step instructions, it might be necessary to have the experimenter stand by near participants' houses in case of hardware troubleshooting. 
    \item[Real-time logging might help monitor task progress.] Visible real-time logging (\eg server logs on the testbeds) can serve as a supplementary information for the sighted experimenter for tracking participants' task progress in addition to Zoom calls.
\end{description}

\subsection{Step 3: Case Study}

\subsubsection{Participants}

A total of 12 blind participants were recruited through emailing lists and local organizations --- our remote user study was reviewed and approved by IRB (\#1255427-6) at the University of Maryland, College Park.
As shown in Table~\ref{tab:participants}, six blind participants self-reported as female and the other six as male. Their ages ranged from 33 to 70 ($Mean=54.3$, $SD=15.2$). Eight participants were totally blind while the other four were legally blind. Five participants (P1, P5, P6, P10, P11) reported having light perception.
As depicted in Figure~\ref{fig:remote_study_design}, blind participants, located in their homes, communicated with the experimenter, located in a car nearby participants' home,  via dual video conferencing. The experimenter monitored participants' activities by having access to the two video streams in the same Zoom call and real-time server logs on a separate window. The first video stream was captured by the camera in the smart glasses with the sound muted, and the second from a laptop camera facing participants. All Zoom sessions were recorded.

\begin{table}[b]
    \centering
    \caption{Demographic information of participants in our study. Asterisks (*) indicate light perception.}
    \begin{tabular}{@{}ccccc@{}}
        \toprule
        PID & Age & Gender & Vision level       & Age of onset \\
        \midrule
        P1  & 39  & Female & Totally blind*     & Birth \\
        P2  & 67  & Male   & Legally blind      & 55 \\
        P3  & 62  & Female & Totally blind      & Birth \\
        P4  & 32  & Male   & Legally blind      & 20 \\
        P5  & 66  & Male   & Totally blind*     & 46 \\
        P6  & 61  & Male   & Totally blind*     & 41 \\
        P7  & 70  & Male   & Legally blind      & Birth \\
        P8  & 50  & Female & Legally blind      & 45 \\
        P9  & 69  & Female & Totally blind      & 55 \\
        P10 & 66  & Female & Totally blind*     & Birth \\
        P11 & 33  & Female & Totally blind*     & Birth \\
        P12 & 36  & Male   & Totally blind      & Birth \\
        \bottomrule
    \end{tabular}
    \label{tab:participants}
\end{table}

\subsubsection{Materials}

\begin{figure}[t]
    \centering
    \begin{subfigure}{0.22\textwidth}
        \centering
        \includegraphics[width=\textwidth]{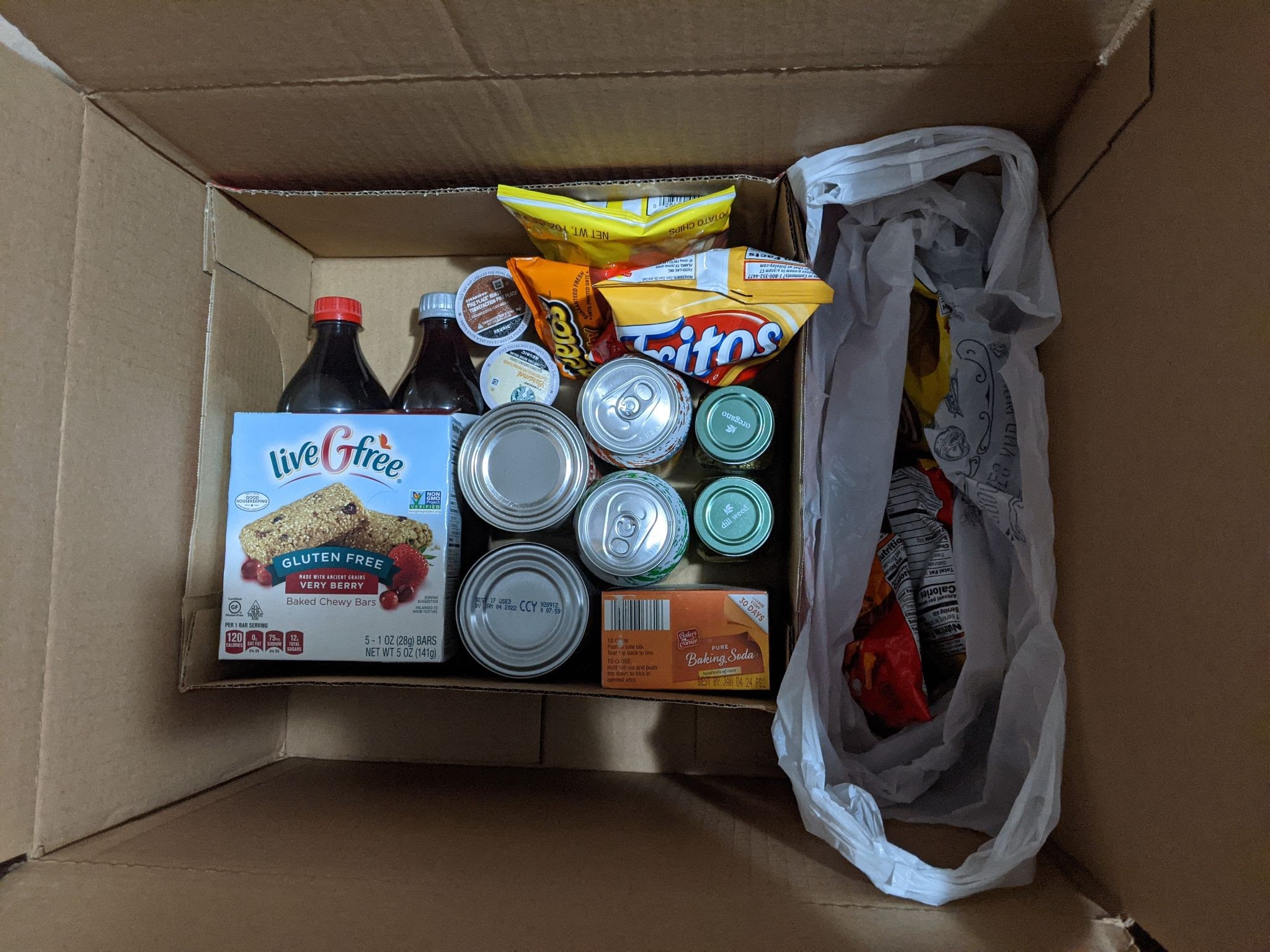}
        \Description[A box contains a small box with 15 stimuli objects and a white plastic bag with 3 snacks.]{A box contains a small box and a white plastic bag. The small box contains a box of baking soda, a caramel coffee k-cup, a pike place roast k-cup, three snacks (Cheetos, Fritos, Lays), a box of chewy bars, a can of chicken broth, a bottle of coca cola, a bottle of diet coke, a can of diced tomatoes, a bottle of dill, a bottle of oregano, a can of lacroix apricot, a can of lacroix mango. The white plastic bag contains three snacks, Cheetos, Fritos, and Lays.}
        \caption{A box including a smaller box with 15 stimuli objects used for the testbed A (left) and a plastic bag with 3 snacks used for the testbed B (right).}
        \label{fig:package_materials}
    \end{subfigure}%
    \quad
    \begin{subfigure}{0.22\textwidth}
        \centering
        \includegraphics[width=\textwidth]{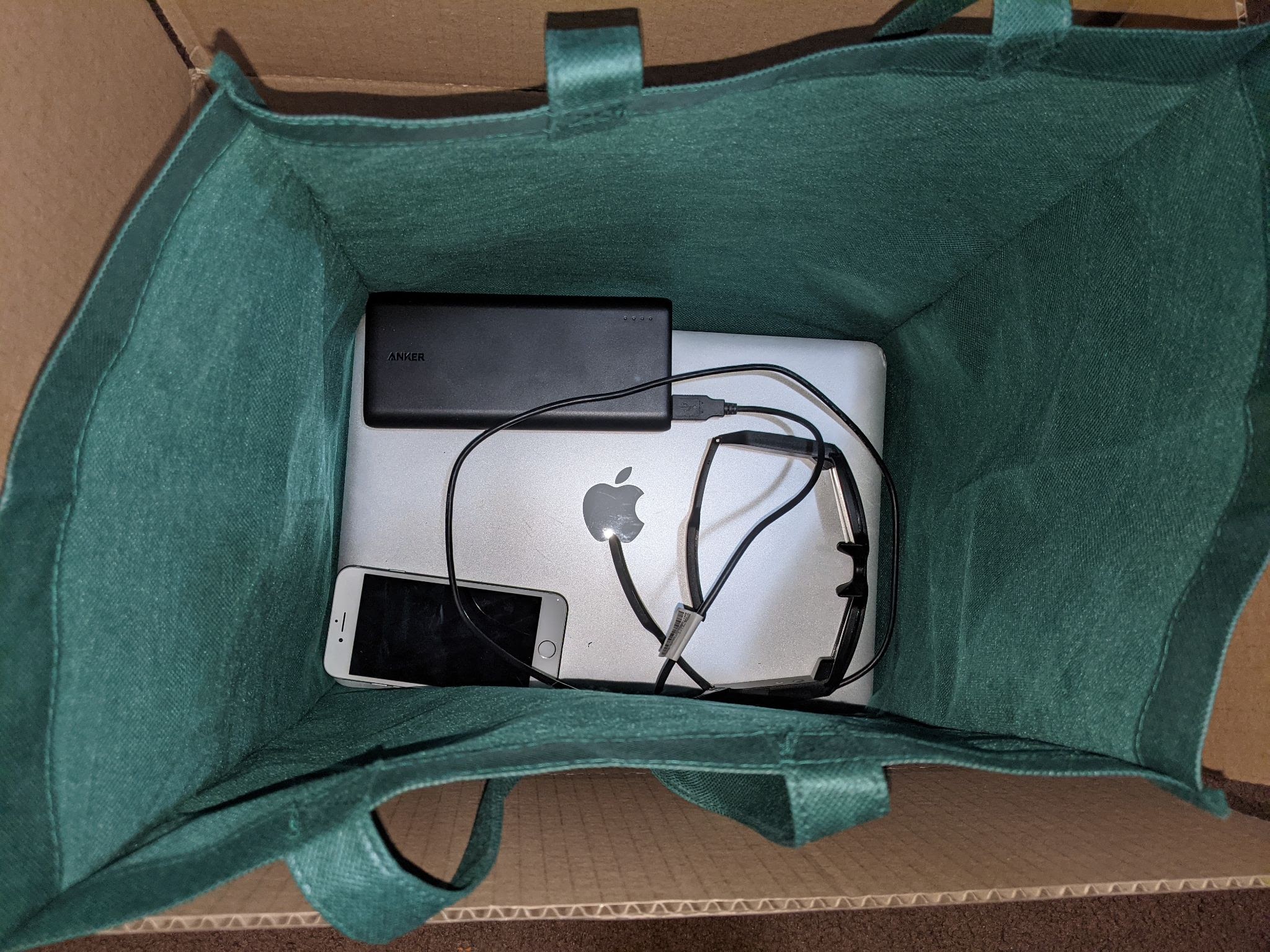}
        \Description[A green tote bag containing hardware devices used in our remote study.]{A green tote bag includes a MacBook, an iPhone, and a portable charger.}
        \caption{A shopping bag with hardware devices used for our remote study: MacBook, smart glasses connected to a portable charger, and iPhone.}
        \label{fig:shoppingbag_materials}
    \end{subfigure}%
    \caption{Two groups of study materials put into two different packages: a box (left) and a reusable shopping tote bag (right). Each package has unique texture. Participants were sent the study materials prior to their study session.}
    \label{fig:study_materials}
\end{figure}

As shown in Figure~\ref{fig:study_materials}, participants received two boxes. The first contained a bag with four devices: a fully charged iPhone with two testbed apps (Testbed A and Testbed B) already installed, a fully charged Macbook with the Zoom call initiated, a pair of fully charged Vuzix Blade glasses connected with an 1.2 meter long cable to a portable charger and an initiated Zoom call, as well as a 5G mobile hotspot device. The second box contained the 15 stimuli objects for Testbed A and three snacks separated with a plastic bag for Testbed B. 

The containers (\ie, the shopping bag, the box, and the plastic bag) were chosen to have different textures distinguishable by touch. Participants were asked to pick up each of these containers at different points during the study. For example, they were first asked to set up the laptop in front of them, wear the smart glasses, and bring out the phone. Then, they interacted with the objects following later instructions about the testbeds, described in Section~\ref{sec:tasks}.

\subsubsection{Environment}
All participants completed the study session remotely from their houses. They were instructed to find a sitting area in which they feel comfortable setting up the laptop and interacting with the stimuli objects. Before starting the session, the experimenter helped participants to orient the laptop camera so that it can include the participants' upper body.

\subsubsection{Tasks}
\label{sec:tasks}
After the initial interview, including questions related to demographics as well as experience and attitudes towards technology, participants performed two sets of tasks.
The tasks involved two different smartphone testbeds for object recognition, which we refer to as \textit{Testbed A} and \textit{Testbed B} respectively in this paper. Testbed A was pretrained by the experimenter to recognize the 15 stimuli objects. Testbed B was trained in real time by the participants to recognize the three snacks that were given to them. 

Each set of tasks started with an onboarding session for the associated testbed and followed by one or four more tasks for Testbed A and B, respectively. Before and after each set, participants answered task-related questions verbally. During the session, participants could freely interact with stimuli objects. For example, some participants placed object stimuli on their table and took photos while others held them in their hands and took photos. Below we include more details on the tasks. However, the participants' feedback on the testbeds and study findings related to the testbeds are beyond the scope of this paper, so we do not include them in our results.

\textbf{Set of tasks for Testbed A}
\begin{itemize}
    \item Onboarding\textsubscript{A}: Pick up objects from the box and find the Testbed A on the smartphone.
    \item Task1\textsubscript{A}: Take photos of 15 stimuli objects using the Testbed A on the smartphone for testing.
\end{itemize}

\textbf{Set of tasks for Testbed B}
\begin{itemize}
    \item Onboarding\textsubscript{B}: Pick up objects from the plastic bag and find the Testbed B on the smartphone.
    \item Task1\textsubscript{B}: Take photos of three snacks using the Testbed B on the smartphone for training.
    \item Task2\textsubscript{B}: Review visual characteristics of the photos as indicated  by the Testbed B on the smartphone.
    \item Task3\textsubscript{B}: Take photos of three snacks using the Testbed B on the smartphone for testing.
    \item Task4\textsubscript{B}: Find specific options on the Testbed B on the smartphone and change them. 
\end{itemize}

\subsection{Step 4: Peer Debriefing and Video Analysis}

After conducting the case study with all 12 blind participants, R2 met with R1 and R5 for peer debriefing to organize findings and observations.
R2 shared his experiences with carrying out the study, focusing on the planning and execution of our study design.

Based on the debriefing analysis, the three researchers created a coding scheme for annotating the Zoom videos recorded by both the laptop camera and the smart glasses. To answer RQ1 \textit{``What do smart glasses worn by blind people capture during a remote user study?''}, they constructed \textbf{Visibility}. To help answer RQ2 \textit{``How can this be leveraged to increase access to their interactions with the system being evaluated and better support experimenter-participant communication?''}, they constructed \textbf{Guidance}. R1 coded video frames from the laptop and smart glasses cameras by following these definitions:

\textbf{Visibility}
\begin{itemize}
    \item \textit{Fully visible}: A set of video frames \textit{fully} show a stimuli object, or a testbed's interface and a participant's interaction.
    \item \textit{Partially visible}: A set of video frames \textit{partially} show a stimuli object, a testbed's interface, or a participant's interaction.
    \item \textit{Not visible}: A set of video frames do \textit{not} show a stimuli object, a testbed's interface, and a participant's interaction.
\end{itemize}

\textbf{{Guidance}}
\begin{itemize}
    \item \textit{Count}: The number of occurrences where the experimenter provides blind participants camera aiming guidance.
\end{itemize}

Zoom videos from the laptop camera and smart glasses were recorded for all participants except for P5 and P8. Videos from P5 and P8 were not annotated since their smart glasses got disconnected from the portable battery in the middle of their study sessions, which led to incomplete video recordings.
During the video annotation, the annotator focused on frames where participants performed specific actions that were necessary for each task on a testbed.
For example, during the onboarding session for Testbed A, the annotator looked at frames where participants interacted with a testbed on a smartphone or picked up stimuli objects from the package. In this example, the annotator checked if the videos captured stimuli objects or the smartphone screen and participants' interaction with it, which are necessary information for the experimenter to check during that task.

\begin{figure}[b]
    \begin{subfigure}{0.47\textwidth}
        \includegraphics[width=\textwidth, left]{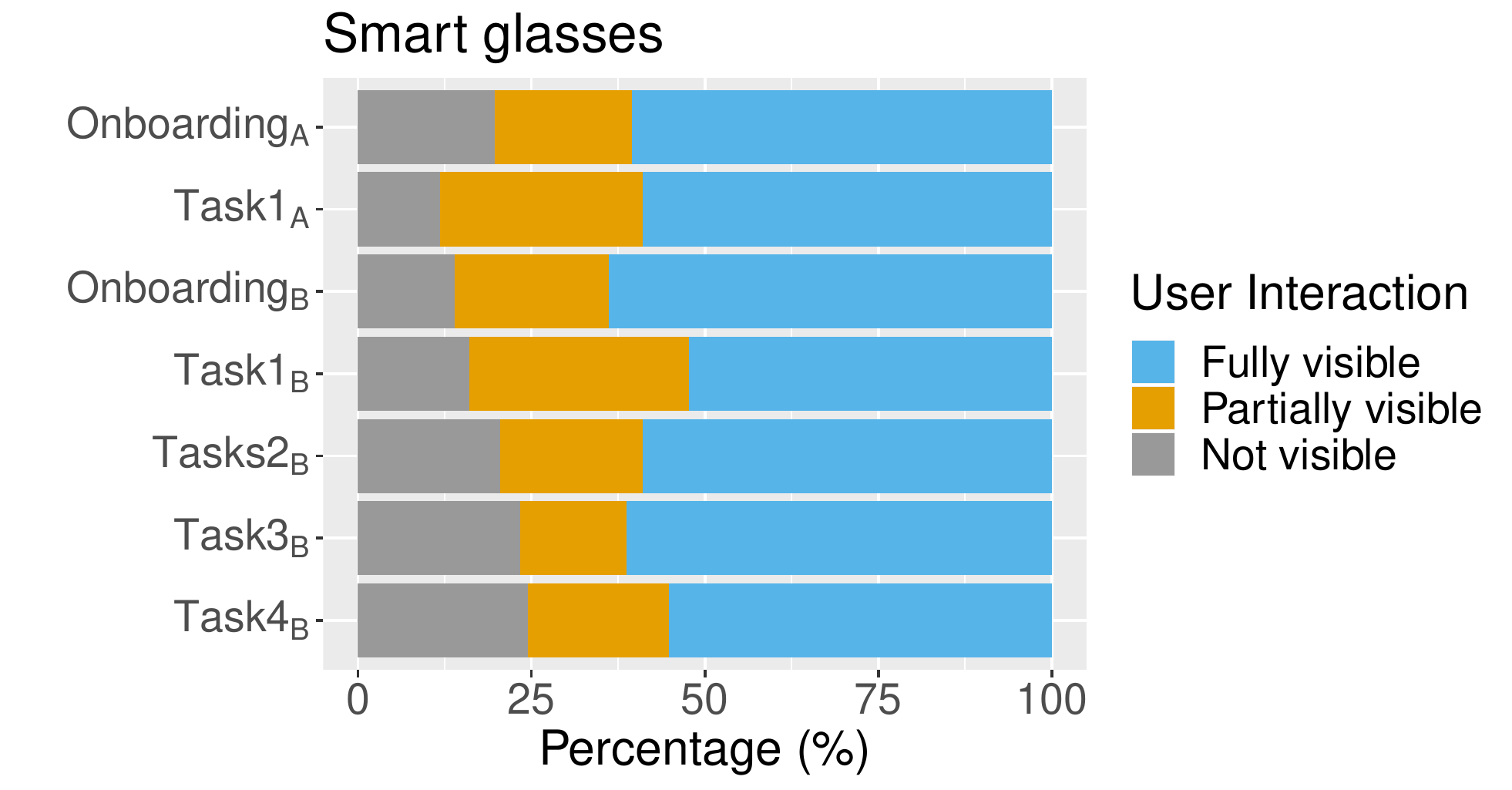}
        \Description[It shows percentages of the fully visible, partially visible, and not visible annotations on Zoom smart glasses videos for seven tasks.]{From top to bottom, there are seven stacked bars for seven tasks, respectively. Here, we present the percentage of each annotation level for each task. TestbedA_Onboarding: fully visible(60.56\%), partially visible(19.72\%), not visible(19.72\%); TestbedA_Task1: fully visible(58.99\%), partially visible(29.21\%), not visible(11.80\%); TestbedB_Onboarding: fully visible(63.89\%), partially visible(22.22\%), not visible(13.89\%); TestbedB_Task1: fully visible(52.30\%), partially visible(31.64\%), not visible(16.06\%); TestbedB_Task2: fully visible(58.97\%), partially visible(20.51\%), not visible(20.51\%); TestbedB_Task3: fully visible(61.29\%), partially visible(15.32\%), not visible(23.39\%); TestbedB_Task4: fully visible(55.21\%), partially visible(20.25\%), not visible(24.54\%)}
        \caption{Annotated visibility from the camera on the smart glasses.}
        \label{fig:vuzix_inclusion}
    \end{subfigure}%
    \vspace{1em}
    \begin{subfigure}{0.47\textwidth}
        \includegraphics[width=\textwidth, left]{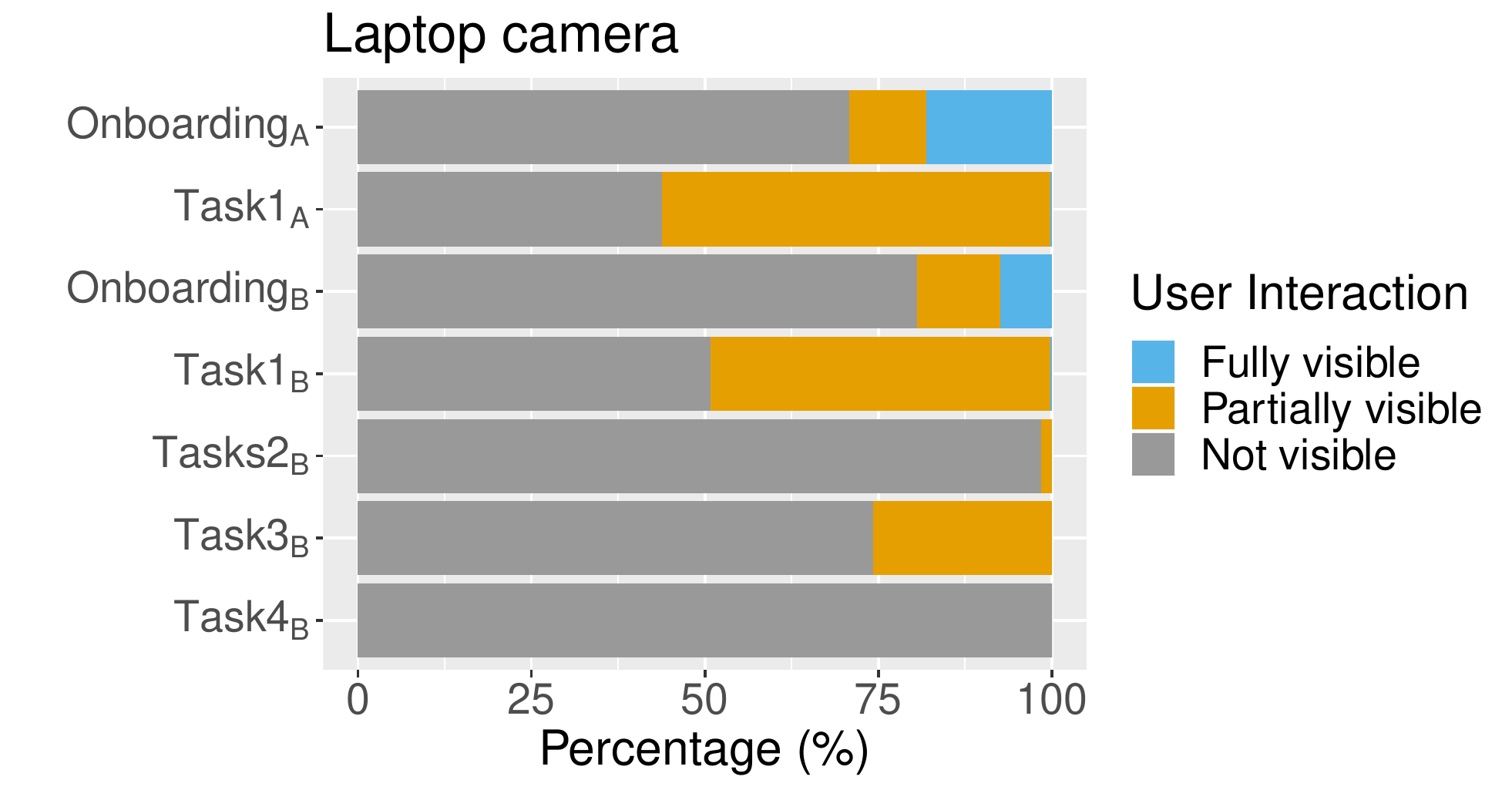}
        \Description[It shows percentages of the fully visible, partially visible, and not visible annotations on Zoom laptop camera videos for seven tasks.]{From top to bottom, there are seven stacked bars for seven tasks, respectively. Here, we present the percentage of each annotation level for each task. TestbedA_Onboarding: fully visible(18.06\%), partially visible(11.11\%), not visible(70.83\%); TestbedA_Task1: fully visible(0.19\%), partially visible(55.96\%), not visible(43.85\%); TestbedB_Onboarding: fully visible(7.41\%), partially visible(12.04\%), not visible(80.56\%); TestbedB_Task1: fully visible(0.16\%), partially visible(49.04\%), not visible(50.80\%); TestbedB_Task2: fully visible(0.0\%), partially visible(1.50\%), not visible(98.50\%); TestbedB_Task3: fully visible(0.0\%), partially visible(25.81\%), not visible(74.19\%); TestbedB_Task4: fully visible(0.0\%), partially visible(0.0\%), not visible(100.0\%)}
        \caption{Annotated visibility from the camera on the laptop.}
        \label{fig:laptop_inclusion}
    \end{subfigure}%
    \caption{A comparison between smart glasses and laptop cameras in terms of percentage of video frames across the tasks where overall participant interactions with the phone and stimuli were fully visible, partially visible, or not visible.}
    \label{fig:inclusion_percentage}
\end{figure}

\section{Findings and Observations}
We enrich our lessons learned during the pilot studies with observations and findings from the peer debriefing and the fine-grained analysis of study recordings, a total of 17 hours and 50 minutes of videos from 10 participants. Specifically, we discuss results on the feasibility of this approach in terms of \textbf{access} (\ie, by looking at the visual information captured by the smart glasses versus the laptop), \textbf{support} (\ie, by looking at the experimenter-participant communication), and \textbf{logistics} (\ie, by reflecting on our experiences with handling delivery and troubleshooting).

\subsection{Access: What is Captured}

\begin{figure*}[t]
    \centering
    \begin{subfigure}{0.4\textwidth}
        \centering
        \includegraphics[width=\textwidth]{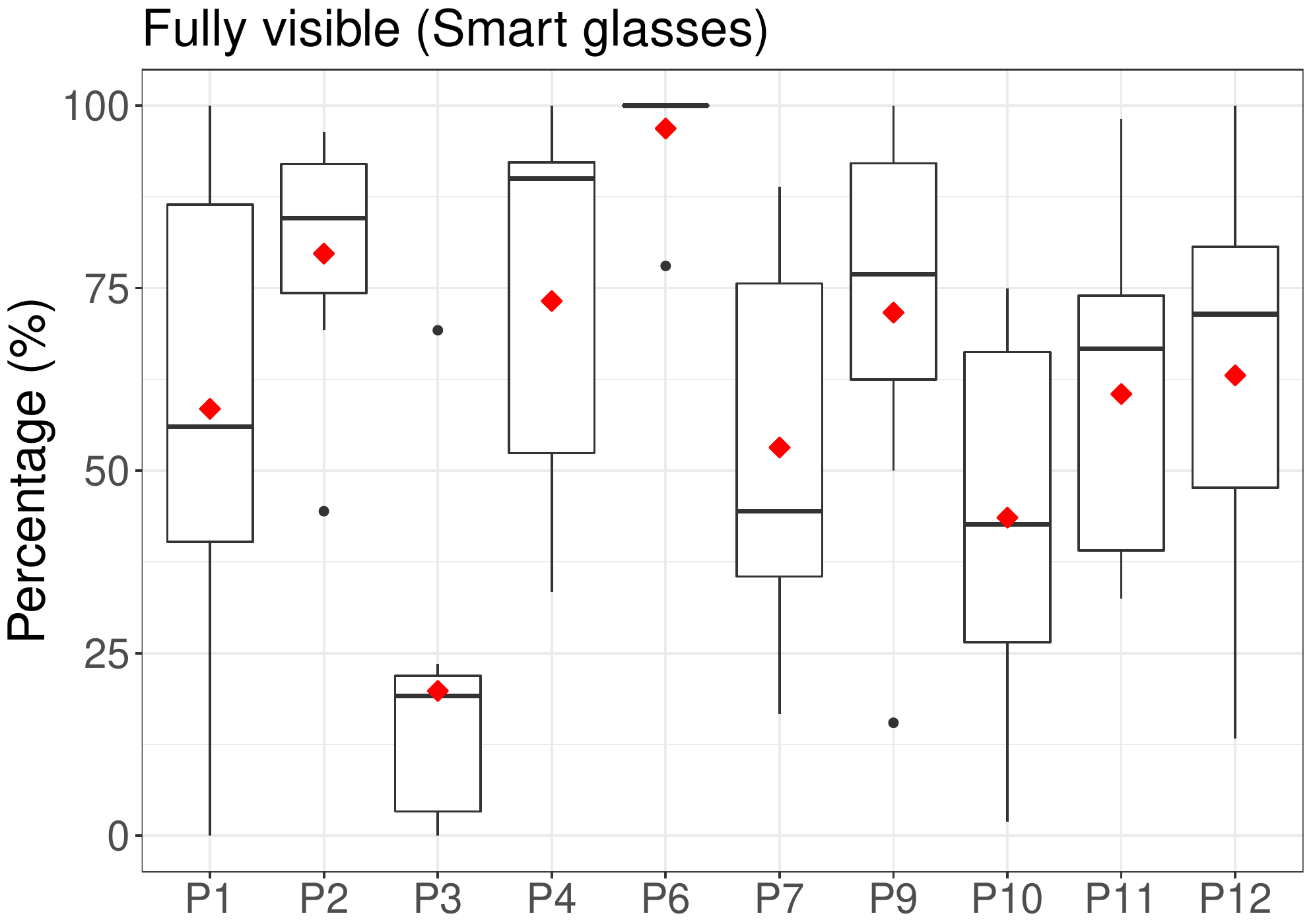}
        \Description[A box plot depicts percentages of the fully visible annotations on each participant's data captured by smart glasses.]{From left to right, there are 10 box plots for 10 participants, respectively. P5 and P8 are not included here. Here, we share the median, mean, and standard deviation (sd) for each box plot. P1: median=56.0\%, mean=58.48\%, sd=36.22\%; P2: median=84.62\%, mean=79.74\%, sd=18.07\%; P3: median=19.16\%, mean=19.83\%, sd=23.80\%; P4: median=90.0\%, mean=73.24\%, sd=28.10\%; P6: median=100.0\%, mean=96.86\%, sd=8.30\%; P7: median=44.44\%, mean=53.18\%, sd=27.59\%; P9: median=76.92\%, mean=71.66\%, sd=30.08\%; P10: median=42.62\%, mean=43.57\%, sd=27.01\%; P11: median=66.67\%, mean=60.50\%, sd=24.47\%; P12: median=71.43\%, mean=63.05\%, sd=29.17\%}
        \caption{Percentage of video frames on the smart glasses of each participant (except P5 and P8) that were annotated as `fully visible' across different tasks.}
        \label{fig:vuzix_fully_visible_per_pid}
    \end{subfigure}%
    \qquad
    \begin{subfigure}{0.4\textwidth}
        \centering
        \includegraphics[width=\textwidth]{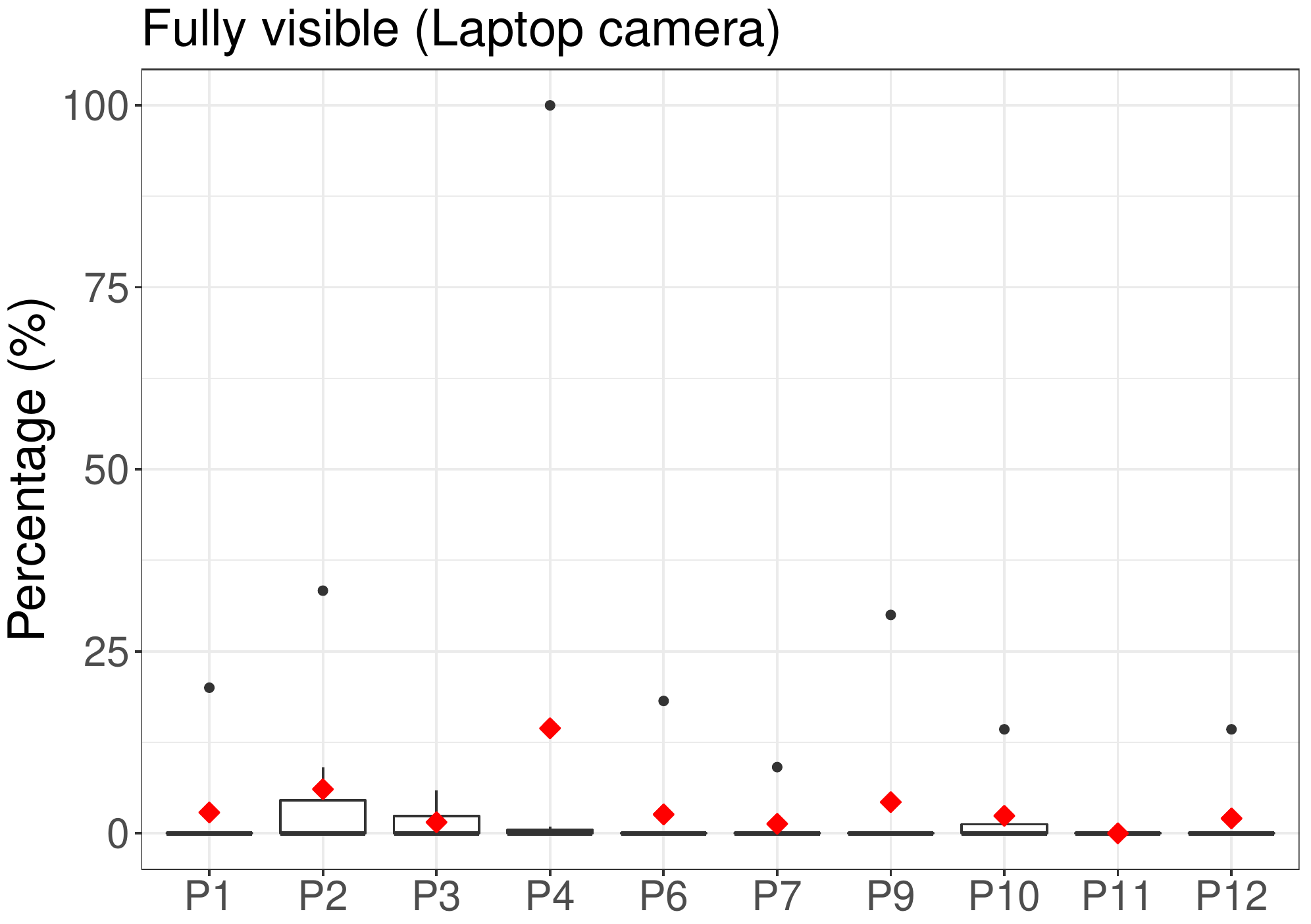}
        \Description[A box plot depicts percentages of the fully visible annotations on each participant's data captured by the laptop camera.]{From left to right, there are 10 box plots for 10 participants, respectively. P5 and P8 are not included here. Here, we share the median, mean, and standard deviation (sd) for each box plot. P1: median=0\%, mean=2.86\%, sd=7.56\%; P2: median=0\%, mean=6.06\%, sd=12.50\%; P3: median=0\%, mean=1.52\%, sd=2.62\%; P4: median=0\%, mean=14.42\%, sd=37.74\%; P6: median=0\%, mean=2.60\%, sd=6.87\%; P7: median=0\%, mean=1.30\%, sd=3.44\%; P9: median=0\%, mean=4.29\%, sd=11.34\%; P10: median=0\%, mean=2.40\%, sd=5.32\%; P11: median=0\%, mean=0\%, sd=0\%; P12: median=0\%, mean=2.04\%, sd=5.40\%}
        \caption{Percentage of video frames on the laptop of each participant (except P5 and P8)  that were annotated as `fully visible' across different tasks.}
        \label{fig:laptop_fully_visible_per_pid}
    \end{subfigure}%
    \caption{A comparison between smart glasses and laptop camera in terms of percentage of video frames across the tasks where each participant interactions with the phone and stimuli were fully visible.}
    \label{fig:fully_visible_per_pid}
\end{figure*}

Focusing on the part of the study where participants' interactions with the testbeds are critical, we find that overall the camera from the smart glasses captures much more information than the static camera from the laptop as can be observed by comparing Figure~\ref{fig:vuzix_inclusion} to Figure~\ref{fig:laptop_inclusion}. This observation seems to be consistent across all tasks that involve interactions with the mobile phone and object stimuli. More so, as shown in Figure~\ref{fig:fully_visible_per_pid}, it is also consistent across all participants\footnote{Please note that the results for P5 and P8 are not included since their video recordings were incomplete and thus were excluded from the analysis.}. 
We find that on average, 58.75\% (sd=3.88\%) of participants' interactions throughout the tasks were \textit{fully visible} via the smart glasses compared to only 3.69\% (sd=6.90\%) via the laptop. However, this gain in visibility was not uniform across participants. For some (\eg, P6) it was as high as 96.86\% on average (sd=8.30\%), and for others (\eg, P3) as low as 19.83\% (sd=23.80\%).

\begin{figure}[t]
    \centering
    \begin{subfigure}{0.45\textwidth}
        \centering
        \includegraphics[width=\textwidth]{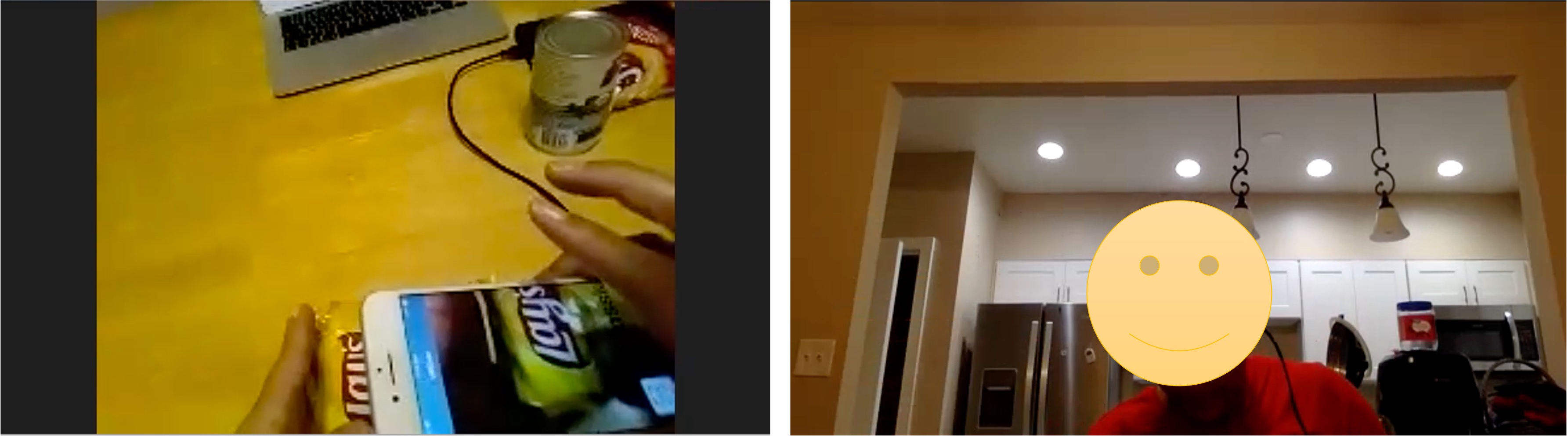}
        \Description[Left: the smart glasses view with the fully visible annotation; right: the laptop camera view with the not visible annotation.]{There are two screenshots. The left one shows that the smart glasses capture a participant's smartphone including a Lays snack. On the other hand, the right one shows that the laptop camera captures only a participant's face and upper body.}
        \caption{An example of a `fully visible' interaction on the smart glasses (left) that was `not visible' on the laptop camera (right).}
        \label{fig:vuzix_fully_visible_example}
    \end{subfigure}%
    \vspace{1em}
    \begin{subfigure}{0.45\textwidth}
        \centering
        \includegraphics[width=\textwidth]{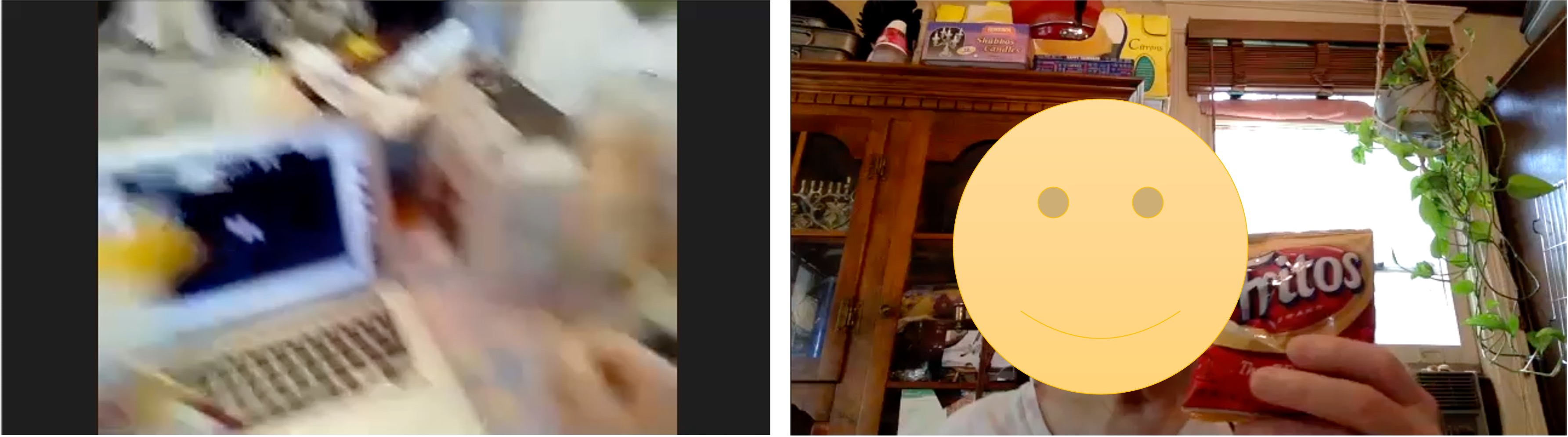}
        \Description[Left: the smart glasses view with the not visible annotation; right: the laptop camera view with the fully visible annotation.]{There are two screenshots. The left one shows that the smart glasses capture a participant's smartphone, not including a Fritos snack. On the other hand, the right one shows that the laptop camera captures a Fritos snack, which is of interest to a participant and an experimenter.}
        \caption{An example of a `not visible' interaction on the smart glasses (left) that was `fully visible' on the laptop camera (right).}
        \label{fig:laptop_fully_visible_example}
    \end{subfigure}%
    \caption{Contrasting examples from the study showing how the two cameras capture complementary information.}
    \label{fig:zoom_examples}
\end{figure}

An interesting observation is that those video frames from the smart glasses, which captured participants interactions with the testbeds, tend to do so in an unobtrusive way as shown in Figure~\ref{fig:vuzix_fully_visible_example}; this resembles more closely an in-lab setup. In contrast, the few video frames from the laptop, where the testbeds or stimuli were fully visible, often resulted from a conscious attempt of the participants to share with the experimenter; as shown in Figure~\ref{fig:laptop_fully_visible_example}, participants had to deviate from a study task and point their phone or stimuli to the laptop camera.  Typically, this occurred at the start of the onboarding tasks, as illustrated in Figure~\ref{fig:laptop_inclusion}. 

Many factors can explain the low visibility of interactions from the laptop camera. First, the camera is static. Thus, participants easily moved out of its field of view. Second, participants were seated during the study and faced the laptop, a typical setup in Zoom calls, which may be appropriate for capturing people's faces but not necessarily their hands, their phones, the objects, or their interactions with them. Asking participants to move the laptop, tilt it, find an ideal framing, restrict their movements, or show every now and then the phone can be disruptive or affect study outcomes.

The camera in the smart glasses overcomes some of these challenges. However, other obstacles remain for accessing interactions via this form factor from all blind participants. Specifically, we observed higher percentages of visibility from the glasses from those participants who became blind later in life compared to those who were born blind. Late blind participants typically aimed their gaze towards the mobile device. Thus, the testbed interface and stimuli were typically included in the smart glasses' camera field of view. However, participants who were congenitally blind often aimed their ears towards the mobile device when interacting with it and anticipating audio feedback (\eg, from the screen reader). One of these participants (P3) maintained this head orientation even when taking photos of object stimuli. As a result, only 19.83\% of P3's video frames on average (sd=23.80\%) were annotated as `fully visible'. Even then, the experimenter was able to leverage many of the video frames from P3's smart glasses, where the interactions were `partially visible'; 35.15\% of the video frames on average (sd=9.44\%). The experimenter triangulated this partial information with the testbed's audio feedback for monitoring study progress.

\subsection{Support: What is Communicated}

Our case study with blind participants demonstrated the potential of smart glasses for interactive communication in a real-time remote setup. The experimenter particularly found smart glasses useful to guide blind participants on picking up stimuli objects in front of them and to observe their actions unobtrusively. More so, observations were mostly done without interruptions (\eg, without asking participants to move stimuli objects or the phone screen towards the laptop camera field of view). Not being obtrusive or interruptive was critical for this particular study, where any priming on object and testbed camera manipulation during tasks that involved taking photos of the objects with the testbed could affect study results. More so, having real-time access to these interactions during participants' photo-taking was important for the experimenter to gain insights into their behaviors and have a better context for the photos that were being collected and analyzed later. 

A unique aspect of deploying smart glasses in studies with blind participants is that audio cues from the testbeds can often help overcome instances of limited visibility. Application serving as testbeds have to be accessible. Thus, they are typically designed to provide audio feedback to participants (\eg, being compatible with screen readers). This audio feedback, which is typically responsive to blind participants actions on the interface, also serves as a good cue for the experimenter. It complements visual cues from the video frames where interactions are 'partially visible' or `not visible'. 

From the annotation of all the videos, we find that only six times the experimenter had to provide guidance for camera aiming; \ie, asking participants to make the stimuli visible to the smart glasses or the laptop camera by moving either their head or their hand slightly. These visibility guidance events mostly occurred at the start of the study in tasks related to the first testbed or during the onboarding for the second testbed.
Only in one out of these six times, the experimenter asked a participant (P2) to show an object to the laptop camera. In all the other times, it was more quickly achieved through guidance for the smart glasses.
Although one participant (P10) struggled with following the guidance to bring a stimuli object into the camera view of the smart glasses, the other participants (P1, P4, P12) promptly reacted to the experimenter's guidance and captured their stimuli objects with the smart glasses --- P1 was given the guidance two times (one during Task1\textsubscript{A} and the other during Onboarding\textsubscript{B}).
In the peer debriefing, the experimenter stated that it was easy to provide camera aiming guidance for smart glasses.

\subsection{Logistics: What is Handled}

\subsubsection{Study equipment delivery}

All the study equipment needed to be delivered to participants and set up before the remote study session. Since our remote method involved expensive, fragile hardware devices (\ie, a smartphone, a laptop, a pair of smart glasses, and a mobile hotspot device), we had to ensure the safe delivery of the study equipment. Instead of relying on a third-party shipping service, one of our team members took responsibility for the equipment delivery. This member sanitized all the study equipment and left it in front of participants' houses. Then, participants picked it up.
The study material was grouped according to its purpose and placed into several different containers with different textures. Grouping the study material with distinguishable containers was effective in communication between the experimenter and participants. More specifically, the experimenter was able to provide participants clear instructions about which material to pick up as it helped participants find the right study material among others.

\subsubsection{Reliable network connection}

Even within the same local area, we observed variance in network latency although the mobile hotspot device deployed in our study supported up to 5G Nationwide network.
Often, such latency made it difficult for participants to communicate with the experimenter and for the experimenter to observe participants' activities remotely in real time.

Moreover, we observed that the mobile hotspot device sometimes overheated due to the overuse of network data. This issue typically emerged when participants took many photos (they were allowed to take as many photos as they wanted in some of the tasks).  These tasks caused a lot of network traffic since the hotspot device needed to send a large number of photos to a remote server and receive real time results from the server at a photo level. All these occurred while the mobile hotspot also supported dual video conferencing on participants' laptop and smart glasses.
The overheated mobile hotspot device first led to increasing network latency and then stopped working; thus, the experimenter had to restart the hotspot device after cooling it down.

\subsubsection{Local troubleshooting}

When a remote user study involves multiple hardware devices that participants may not use in their daily lives, troubleshooting is unavoidable. Although we noticed that blind participants learned how to put on smart glasses quickly, local troubleshooting of hardware devices, including a Macbook, an iPhone, and Vuzix Blade smart glasses, was necessary for some participants. For example, the experimenter found remote instructions insufficient to help some participants address unexpected software issues, especially when they were not familiar with default settings on the laptop or mobile device --- some of them were used to their customized configurations.
One challenge with dual video conferencing (laptop and smart glasses that are co-located) is that it can lead to voice echoing when both of the mics are on. This happened in few cases when participants accidentally activated the mic on their smart glasses.
The experimenter managed to mute one of the mics to address the issue, but noticed that there was no way to unmute any when both of the mics were accidentally muted. It occurred once to a participant who was not familiar with Zoom; thus, the experimenter had to go and fix this issue locally.
Furthermore, hardware issues, such as cable disconnection or hardware malfunctions, were typically inaccessible for blind participants to spot. Thus, when participants faced such hardware-related issues, the experimenter had to fix the issues locally after retrieving the hardware device from them. In our case study, the experimenter was standing by near participants' house while remotely conducting a user study. However, we suggest having at least two experimenters during a remote user study session --- one conducting the study remotely and the other troubleshooting device issues locally.

\section{Discussion}
In this section, we first reflect on lessons learned from accessing blind participants’ interactions via smart glasses in remote studies, while discussing the implications for designing inclusive smart glasses and employing this technology in remote studies.  We then discuss limitations in our study that may affect the generalizability of our findings as well as future work.

\subsection{Implications}
Our study and findings provide evidence for the potential of smart glasses in remote usability testing with blind participants. We see how researchers who plan remote studies with this population and those who are interested in designing more inclusive smart glasses could benefit from the following insights:

\begin{description}[leftmargin=0.3cm]
    \item [Increasing remote access.] Video conferencing via smart glasses can provide real-time remote access to blind users’ interactions with a mobile application and stimuli from a first-person perspective. Our analysis indicates that this approach surpasses typical video conferencing setups employing a laptop camera. Specifically, we find a striking difference between participants' interactions being fully visible via the smart glasses (on average, 58.7\%) and the laptop (on average, 3.7\%). Even partially visible interactions to the smart glasses can help the experimenter to triangulate participants' interactions with audio cues such as screen reader output indicating users’ actions.
    \item [Supporting real-time communication.] Video conferencing via smart glasses can support real-time experimenter-participant communication. Our analysis indicate that the video stream from the smart glasses supported the experimenter in providing guidance related to study tasks and making observations in an unobtrusive way. Interruptions related to camera aiming were minimal and only occurred for a few participants usually during onboarding; for almost all of the visibility was more quickly achieved via the smart glasses than the laptop. 
    \item [Exploring camera field of view for inclusion.] What is captured by the smart glasses may relate to the age at onset of blindness as camera viewpoint is susceptible to head movements, especially when the camera field of view is limited. Our analysis of video frames from Vizux Blade, which has a 64-degree horizontal field of view, indicates a higher percentage of visibility from the glasses for participants who became blind later in life compared to those who were born blind and tend to ``turn their head to orient their ear towards a sound source''~\cite{wanet1985processing, martinez1977informations, lewald2002opposing}. This observation provides additional evidence on prior work indicating how limited field of view in smart glasses for pedestrian detection can exclude blind users who could benefit the most from this technology~\cite{lee2021accessing}. While smart glasses with a wider camera angle could potentially help, they can also lead to image distortions~\cite{stearns2018automated}. Our study highlights the need to explore the effect of camera field of view on the inclusion of blind users, especially those with early onset who may exhibit distinct head movements from sighted people.
    \item [Prioritizing screen reader support for smart glasses.] It is surprising to see how even smart glasses that run on operating systems supporting accessibility do not include many of those features. For example, accessibility on Vuzix Blade, the smart glasses in our study, is very limited.  Vuzix Blade is an Android device, but it does not support TalkBack, the Google screen reader that is typically included on Android devices. Even smart glasses that are specifically designed for people with visual impairments often opt for ``hands-free'' interactions via voice command support rather than providing a full screen reader experience \eg, on touch interactions. While there are some workarounds that can be employed in remote studies (\eg, by setting up the Zoom call aforetime), unexpected challenges may still arise.  For example, participants can accidentally activate or deactivate the mic on the Zoom call on their smart glasses by touch. As with the early Web, accessibility in wearable devices appears to be an afterthought. It cannot continue this way and be effective. Inclusive design process is essential to make wearable technology effective and accessible for all.
    \item [Overcoming the rapidly draining battery.] Power remains one of the main challenges in smart glasses.  This drawback of limited battery is exacerbated in remote studies where smart glasses are used for video conferencing. Even workarounds like long power cables and portable chargers can fail as cable disconnection or hardware malfunctions can be inaccessible for blind participants to spot; \eg, it happened to two participants in our case study. This unpredictability suggests that smart glasses should be used as a complement to the typical call setup (\eg, video call on a laptop or a phone call) between the experimenter and participant in remote studies.
\end{description}

We see how some of the insights above, which are directly tied to our smart glasses approach for accessing blind participants' interactions with a mobile device and object stimuli, may be adapted for other \textit{testbeds} (\eg smartwatch applications), \textit{settings} (\eg when in movement), or \textit{populations} (\eg studies where accessing head and gaze information is critical or when the first-person perspective can provide more context). More so, our pilot sessions and case study with blind participants may offer more practical insights into the challenges and logistics for those planning to conduct their studies remotely, independently of whether they employ smart glasses or not. For example, lessons learned relate to (i) use of different textures for helping participants quickly distinguish study materials, (ii) use of real-time logging for monitoring task progress when screen recording is not an option, (iii) use of multiple internet hotspots separating video streaming from testbed network traffic, (iv) nearby presence of an experimenter in addition to a remote experimenter, and (v) duplicating some of the study equipment to reduce risks associated with dependency on equipment delivery. 

\subsection{Limitations and Future Work}
There are many limitations that could impact the generalizability of our findings. Our observations come from a single case study (as well as pilot sessions) conducted by one experimenter, on a single evaluation task (mobile applications for object recognition), in a given area within one country (Maryland, the United States). 

The participant pool was small ($N$=12) although it is typical for a user study in accessibility~\cite{mack2021what} and human-computer interaction~\cite{caine2016local}, balanced in terms of male and female participants, and somewhat diverse in terms of age, vision level, and age of onset.

More importantly, insights we obtained from the case study are limited as they were solely derived from peer debriefing among researchers and analysis of video recordings from our blind participants. It does not include explicit input from blind participants in terms of their experiences with the smart glasses or the overall remote study and how their remote experiences may differ from any prior studies they may have attended in-person in research labs. The case study itself, with the goal of understanding blind participants' experiences with particular testbeds (beyond the scope of this work), already took almost 2 hours. Thus, the researchers opted not to ask participants any further questions, meaning that this work does not include potential challenges that may arise from deploying smart glasses in the houses of blind participants. In contrast to a static laptop camera whose viewpoint could be fixed to a specific area in participants' houses, the viewpoint of the smart glasses is dynamic and dictated by their head movements. The smart glasses may accidentally capture scenes and information that the blind participant may not feel comfortable sharing. Investigating potential privacy risks of deploying smart glasses or any always-on camera in blind people's houses is a topic that we believe is critical to explore in the future.

\section{Conclusion}
In this work, we examined the feasibility and challenges of using smart glasses for a user study that, due to the pandemic, had to move from a lab to blind participants' houses. 
Taking an iterative approach, we devised a remote experimental setup and protocol, in which an experimenter can observe a participant's interactions with smartphone testbeds and object stimuli via dual video conferencing: (i) a laptop camera facing the participant and (ii) smart glasses worn by the participant. We shared our findings, observations, and lessons learned from five pilot sessions and a case study with 12 blind participants.  Specifically, we found that smart glasses could help the experimenter view participants' interactions with a testbed and allow the experimenter to communicate with participants without asking them to deviate from their tasks. We observed that there was a difference between video streams captured by late blind and congenitally blind participants; smart glasses worn by those who became blind later in life tend to capture their interactions more often than smart glasses worn by those who born blind. These observations seem to echo prior work indicating that smart glasses with a narrow field of view can be more susceptible to differences in head movements such as directing one's ear, instead of eye gaze, towards the source of sound. 
Last, we shared our experiences with attempting to overcome challenges in conducting a remote user study with blind participants, such as lack of screen reader support on the smart glasses, a limited battery of the smart glasses, and ensuring Internet connectivity from participants' houses.

\begin{acks}
We thank the anonymous reviewers for their constructive feedback on an earlier version of this paper. This work is funded in part by NIDILRR (\#90REGE0008) and NSF (\#1816380). The  opinions  and results  herein  are  those  of the  authors  and  not necessarily  those  of the  funding  agencies. 
\end{acks}

\bibliographystyle{ACM-Reference-Format}
\bibliography{references}



\end{document}